 \documentclass[11pt]{amsart}

\usepackage{ulem}
\usepackage{amssymb}
\usepackage{graphicx}
\usepackage{bbm}

\theoremstyle{definition}

\theoremstyle{remark}

\newtheoremstyle{quest}{\topsep}{\topsep}{}{}{\bfseries}{}{ }{\thmname{#1}\thmnote{ #3}.}
\theoremstyle{quest}

\newcommand{\N}{\mathbb{N}} % natural numbers.
\newcommand{\midskip}{\smallskip \smallskip \smallskip}

\begin{document}

\title{Cascades \& Rating Games}

\author{Oussama Fadil, Jake Soloff}
%\address{Brown University, Providence, RI 02912}
%\email{Jake\_Soloff@brown.edu}

%\begin{abstract}
%Abstract Goes HERE
%\end{abstract}

\maketitle

\noindent \textbf{Introduction.}
The current study is aimed at understanding the mechanisms governing online rating systems such as Yelp, IMDB etc. The so-called referral systems have become so popular over the years one can only wonder how they work and if they do at all. \midskip

\noindent Online ratings are based on users' impressions and opinions they form after using a certain product or service. Hence, the first unknown in the equation is the mechanism under which a user decides to purchase a certain good or use some service. We identify two factors that come into play in such a choice; a subjective factor and an objective factor. 
The subjective factor is the impression the user forms on the product or business; the gut feeling that tells him whether it is worth his time and money. The gut feeling is not always reliable; how many times have you had a good impression of a restaurant only to find out that it is not entirely to your taste? The reliability of the customer's gut feeling turns out to be an important factor in the model. 
The objective factor relates to the quality of the business assessed independently of users' tastes or preferences. Often times, the online rating of a business is taken as a proxy for its objective ?quality rating?. The underlying assumption is that by averaging ratings across users of varying tastes and preferences, dependencies upon said factors are eliminated. \midskip

\noindent It is essential to note the importance of subjectivity in a user's choice. We stress the fact that a business or product is not simply good or bad; rather, it is good or bad from a certain user's perspective. Businesses that are perceived as good by a majority of users are labeled as ``high quality" businesses and businesses that are perceived as good by a minority of users are labeled as ``low quality" businesses. A natural question to ask is whether inherently high-quality businesses can ever be labeled as low quality through online rating systems. \midskip

\noindent Next we examine the mechanism under which online ratings are formed. A user chooses to try a certain business or not by a rational evaluation of the subjective and objective factors under a user-agnostic utility function. Upon trying a business the user leaves a rating that is a reflection of his own subjective perception of its quality. Given that we take the online rating to be the objective factor in the decision making process, a user's action is affected by the actions of all the users who have tried the business in the past, hence the notion of a sequential move game.  \midskip

\noindent Under this model, it is clear that the ?objective? online rating is sequentially updated through users' subjective ratings. Thus at its early stages, a business is at the mercy of the subjectivity of its customers. In theory, some businesses could die simply because they were first approached by the wrong customers; others could be overrated by similar mechanisms. In the long run however, one would expect the online rating to converge to an objective measure of a business?s quality. We seek to determine how often and under what conditions this is the case. \midskip

\noindent\textbf{Notation.} We write 
$$X\sim \text{Bern}(p) \text{ or } X\sim \text{Unif}_{[0,1]}$$
to express that a random variable $X$ is distributed according to a Bernoulli distribution with parameter $p\in [0,1]$ or according to a uniform distribution over the interval $[0,1]$, respectively. Furthermore if $P$ is a property we write $$\mathbbm{1}\left\{P\right\} = \left\{\begin{array}{lr} 1 \text{ if } P \text{ holds,}\\
0 \text{ otherwise.}
\end{array} \right.$$

\noindent With this notation in mind, we are now equipped to introduce more formally the general format of what we call a \textit{rating game}. \midskip

\noindent \textbf{Rating Game.} Suppose consumers $i\in \N = \{1,2,3,\dots\}$ must make some decision sequentially according their order. The decision might be whether to adopt a product or service of a firm whose overall quality is given by $\alpha\in [0,1]$. In our consideration we take the firm to be a restaurant, and the restaurant quality $\alpha$ is taken to represent the portion of the total population who will enjoy their experience at the restaurant (i.e. the objective factor). Each consumer $i$ will independently either like or dislike the restaurant according to $v_i \sim \text{Bern}(\alpha)$, but $i$ will only learn of this value if she goes to the restaurant. Instead, consumer $i$ knows the value of a $\textit{private signal}$ $x_i$ (the subjective factor), which is equal to $v_i$ with probability $\rho$. That is, 
$$x_i\sim \text{Bern}(\rho v_i + (1-\rho)(1-v_i)),$$
so if $v_i = 1$ then $x_i\sim \text{Bern}(\rho)$ is accurate to $v_i$ with probability $\rho$; and if $v_i = 0$ then $x_i\sim \text{Bern}(1-\rho)$ is also accurate to $v_i$ with probability $\rho$. In a sense, then, $\rho$ represents how well consumers know what they like. The value of $\rho$ is common knowledge, so we may assume that the private signals are on average accurate at least half the time $\rho\geq 0.5$ (indeed if $\rho < 0.5$ then each consumer takes her new private signal to be $1-x_i$, which is accurate to $v_i$ on average $1-\rho > 0.5$ of the time). \midskip

\noindent Each consumer $i\in\N$ must decide in order whether to attend the restaurant, and she knows (1) her own private signal $x_i$ and (2) the average value of the $v_j$'s for consumers $j<i$ who have attended the restaurant already. If $N_i$ is the set of all $j < i$ who attended the restaurant, we denote this average
$$\bar v_i := \frac{1}{|N_i|}\sum_{j\in N_i} v_j,$$
and when $N_i$ is empty we take $\bar v_i = \frac{1}{2}$ (this \textit{initialization} is an important assumption, and we consider later generalizations of the rating game where the restaurant can choose the initialization). We call the value $\bar v_i$ the \textit{rating}. As a summary of this game, we have
\begin{itemize}
\item Players -- consumers $i\in \N$.
\item Moves -- each $i\in \N$ decides, in order, whether to attend a restaurant.
\item Information -- each $i\in \N$ knows her private signal $x_i$ and the current rating $\bar v_i$.
\item Payoffs -- if $i$ does not attend the restaurant, her payoff is $0$; otherwise, her payoff is $2(v_i-\frac{1}{2})$, i.e. if she does like it, her payoff is $1$ and if she doesn't, her payoff is $-1$. \midskip
\end{itemize}

\noindent\textbf{Solution.} Individuals do not observe their type. Hence, in order to decide whether to go to the restaurant or not, each individual will proceed by maximizing over types his ex-ante expected payoff given some signal $x_i$. Recall that each individual has a type $v_i\in\{0,1\}$, observes a private signal $x_i\in\{0,1\}$ and choose an action $a_i\in\{0,1\}$, where $a_i=1$ indicates that he goes to the restaurant and $a_i=0$ indicates that he does not. The ex-ante expected payoff from choosing action $a_i$ for individual $i$ is denoted as $\mathbb{E}\big[u(a_i,v_i)|x_i\big]$, where $ u(a_i,v_i)$ is the utility individual $i$ gets by choosing action $a_i$ when his type is $v_i$. The optimal choice of action for individual $i$, denoted $a_i^*$, is given by:
\small

\begin{align*}
a_i^*
&=\underset{a_i}{\operatorname{argmax}}\ \mathbb{E}\big[u(a_i,v_i)|x_i\big]\\
&=\underset{a_i}{\operatorname{argmax}}\sum_{q\in\{ 0,1 \} } u(a_i,q )p(v_i=q|x_i) \\
&=\underset{a_i}{\operatorname{argmax}}\ \frac{1}{p(x_i)}\sum_{q\in\{ 0,1 \} } u(a_i,q )p(v_i=q,x_i) \\
&=\underset{a_i}{\operatorname{argmax}}\ \sum_{q\in\{ 0,1 \} } u(a_i,q )p(v_i = q)p(x_i | v_i=q) \\
&= \left\{\begin{array}{lr} \mathbbm{1}\left\{\sum_{q\in\{ 0,1 \} } u(1,q )p(v_i = q)p(x_i=1 | v_i=q) > \sum_{q\in\{ 0,1 \} } u(0,q )p(v_i = q)p(x_i = 1| v_i=q)\right\} \text{ if } x_i = 1 \\
\mathbbm{1}\left\{\sum_{q\in\{ 0,1 \} } u(1,q )p(v_i = q)p(x_i=0 | v_i=q) > \sum_{q\in\{ 0,1 \} } u(0,q )p(v_i = q)p(x_i = 0| v_i=q)\right\} \text{ otherwise.}
\end{array} \right. \\ 
%&= \left\{\begin{array}{lr} 1 \text{ if } \sum_{q\in\{ 0,1 \} } u(1,q )p(v_i = q)p(x_i | v_i=q) > \sum_{q\in\{ 0,1 \} } u(0,q )p(v_i = q)p(x_i | v_i=q) \\ 0 \text{ otherwise.} \end{array} \right \\
&=  \left\{\begin{array}{lr}
\mathbbm{1}\left\{-(1-\alpha)(1-\rho)+\alpha\rho>0\times (1-\alpha)(1-\rho)+0\times \alpha\rho\right\}\text{ if }x_i =1
\\
\mathbbm{1}\left\{-(1-\alpha)\rho+\alpha(1-\rho)>0\times (1-\alpha)\rho+0\times \alpha(1-\rho)\right\}\text{ if }x_i =0
\end{array} \right. \\
&= \left\{\begin{array}{lr}
\mathbbm{1}\left\{-1+\alpha+\rho-\alpha\rho+\alpha\rho>0\right\}\text{ if }x_i =1
\\
\mathbbm{1}\left\{-\rho+\alpha\rho+\alpha-\alpha\rho>0\right\}\text{ if }x_i =0
\end{array}\right. \\
&=\left\{\begin{array}{lr}
\mathbbm{1}\left\{\alpha>1-\rho\right\}\text{ if }x_i =1
\\
\mathbbm{1}\left\{\alpha>\rho\right\}\text{ if }x_i =0
\end{array}\right. 
\end{align*}

\normalsize

\noindent More generally, the derivation of the optimal action choice doesn't require an explicit functional form for the utility function $u(a_i,v_i)$. The only assumption made is that $u(a_i,0) + u(a_i,1) = 0$. In other words, the payoff from going to a low quality restaurant is compensated by the payoff from going to a high quality restaurant. Similarly, the payoff from not going to a low quality restaurant is compensated by the payoff from not going to a high quality restaurant. \midskip

\noindent On the other hand, the optimal actions do make a lot of sense. First, individuals require more compelling evidence to go to a restaurant when they receive a negative private signal $\left(a_i^*=1\text{ when }x_i =0\text{ if }\alpha>\rho\right)$ as opposed to a positive private signal $\left(a_i^*=1\text{ when }x_i =1\text{ if }\alpha>1-\rho\right)$. Moreover, we note that the ``objective quality parameter" $\alpha$ plays the role of an adjustment. Naturally, the private signal $x_i$ is noisy and doesn't always coincide with $v_i$. As a result, the individual doesn't simply follow the private signal $x_i$ but factors in the information carried by $\alpha$. Precisely, if individual $i$ is more likely to enjoy the restaurant than his signal is to be right $\left(\alpha > \rho\right)$, the optimal action specifies that he should go to the restaurant regardless of his private signal $a_i^* = 1\ \left(\forall\ x_i\in\{0,1\}\right)$. Intuitively, if $\left(\alpha > \rho > 0.5\right)$ and $x_i=1$, then everything indicates that he will enjoy the venue and the rational choice can only be to dine there. If on the other hand $\left(\alpha > \rho\right)$ and $x_i=0$, then he must choose whether to follow his private signal or the objective rating. Given that the private signal is more likely to be wrong than he is to not enjoy the venue, the rational choice is to dine at the venue. By analogy, if the individual is less likely to enjoy the restaurant than his signal is to be wrong then even with a positive private signal ($x_i=0$), the individual chooses to avoid the venue $\left(a_i^*=0\right)$. By contrast, when the objective quality parameter doesn't fall in any of these two edge cases $\left(\rho>\alpha>1-\rho\right)$, the individual simply follows his private signal: 
\begin{align*}
a_i^*&=\left\{\begin{array}{lr}
\mathbbm{1}\left\{\alpha>1-\rho\right\} = 1\text{ if }x_i =1
\\
\mathbbm{1}\left\{\alpha>\rho\right\}=0\text{ if }x_i =0
\end{array}\right.
\iff a_i^* = x_i
\end{align*}
    Thus, the noisy signal $x_i$ is trusted unless there is compelling evidence from $\alpha$ that sways the balance of probabilities in one direction or the other. \midskip
    
\noindent Finally, we note that individuals do not observe the objective quality parameter $\alpha$ and are forced to estimate it somehow. The essence of our work lies in the fact that individuals are assumed to use the average online rating $\bar{v_i}$ as an estimator of $\alpha$. The fact that $\bar{v_i}$ is not necessarily a consistent estimator of $\alpha$ is not an issue; it is precisely what makes the results interesting. \midskip
    
\noindent\textbf{Comparison.} We consider the similarities and differences with the \textit{rating game} and other games in which individuals must combine noisy private information with inferred information from other players. A motivating scenario from probability is usually described as follows: \midskip

\noindent A prison warden has been given explicit directions to reveal to three prisoners on death row---\textbf{A}lice, \textbf{B}ob, and \textbf{C}harlie---precisely the information that one of them (chosen at random) had been pardoned and that the remaining two of them will be executed in the coming week. None of the prisoners can communicate to each other, but they each speak with the warden regularly. Prisoner \textbf{A} asks the warden to reveal who will be executed. The warden, careful not to reveal to \textbf{A} any information on her own fate, reveals that \textbf{B} will be executed. \midskip

\noindent The point is that \textbf{A} may mistakenly reason that, now since one of \textbf{A} or \textbf{C} must be executed, her chances of being executed have risen from $\frac{1}{3}$ to $\frac{1}{2}$. The warden was clever to realize, however, that the probability of \textbf{A}'s execution is independent of the news of \textbf{B}'s execution, so the probability remains at $\frac{1}{3}$, which one shows using Bayes' rule. A similar tradeoff between having tangible information and being able to make use of it is realized gradually in the \textit{rating game}. At first, player $1$ has no information of other players' preferences and must make her decision based solely on her private signal $x_1$ and the initial rating $\bar v_1 = \frac{1}{2}$; suppose player $1$ does visit the restaurant, in which case player $2$ knows $x_2$ and $\bar v_2 = v_1$. So player $2$ knows exactly whether or not player $1$ liked the restaurant, but this tells him very little more about his own preferences. Since $v_i$ varies from person to person, player $2$ wants a more \textit{aggregate} indication of how well consumers liked the restaurant. As $i\in \N$ increases and more players have contributed to the rating, those who have decided to try the restaurant may have become a self-selective pool, relying both on their private signal and the current rating. So while $\bar v_i$ for small $i$ depends greatly on the inclinations of a few consumers, $\bar v_i$ as $i$ gets larger may become a substantially biased estimator of $\alpha$; as we shall see, it depends in significant ways on how well individuals know their own tastes. \midskip

\noindent For example, in the case where $\rho = 1$, all the consumers know precisely whether they like the restaurant because $x_i = v_i$. So consumers do not need to know the rating in order to decide: they just follow their private signals. It is interesting to note, however, that in this special case we have the rating converges $\bar v_i \rightarrow 1 = \rho$ as $i\rightarrow \infty$ for any $\alpha > 0$. \midskip

\noindent An important related model depicts an \textit{information cascade}. In one example, an employer might find a job applicant very promising, but knowing that the applicant has been rejected for similar jobs in the past may cause the employer to ignore his own impression and reject the applicant anyway; in another example, a stock value can become overinflated when individuals ignore their personal evaluations of its quality because they see the stock value spike. In this model the users all get the same value $v_i = v\in\{0,1\}$ from the restaurant, so we might think of a rating game with $\alpha\in\{0,1\}$ chosen according to a $\text{Bern}(\frac{1}{2})$. That way $v_i = \alpha$ for every consumer $i\in\N$ but they simply don't know whether the restaurant is good or bad. Instead of getting a rating from previous consumers, in an information cascade model, consumer $i$ knows whether each previous $j < i$ decided to try the restaurant. Each consumer $i$ also has their private signal $x_i$ which still depends on $\rho$ in the same way. Supposing that indeed $v_i = \alpha = 1$ (individuals will like the restaurant if they go) we showed (in an assignment) that in this model, by the time individual $3$ makes her decision, already with probability $\rho^2$ is everyone willing to ignore their private signal and go to the restaurant (a `correct cascade') and with probability $(1-\rho)^2$ is everyone willing to ignore their private signal and avoid the restaurant (an `incorrect cascade'). We made two essential alterations when coming up with the \textit{rating game}: $v_i$ depends on the individual $i$ in our model, and each individual cannot see the decisions of the previous, only the rating they leave. Both changes are made partially with the hope that it becomes more feasible to prevent either cascade and partially as a nice context to study the rating system. \midskip

\noindent {\textbf{Experiments and Results.}} Fixing the strategy described in the `solution' section, it begins to make sense to ask about simulating this game, since some quantities like the rating $\bar v_i$ converge as $i$ gets very large. One quantity we might study using simulated data is the average rating, to see if it converges to what we expect. Fixing $\rho = 0.8$, we plot the average rating $\bar v_i$ versus $i$ for various $\alpha$. 

\begin{figure}[h!]
  \centering
    \includegraphics[width=0.7\textwidth]{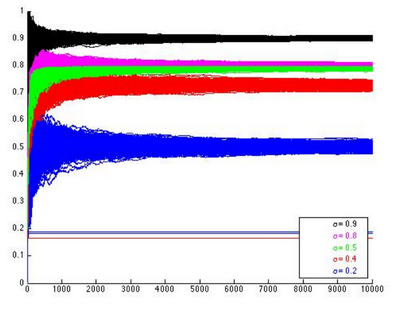}
  \caption{Rating $\bar v_i$ versus consumers $i$ for various $\alpha$.}
\end{figure}

\noindent When $\alpha > \rho$ we have $\bar v_i\rightarrow \alpha$ as we expected. We interpret this to mean that whenever enough people will like the restaurant, it does not matter how good people are at knowing their own preferences because a critical mass of people who try the restaurant and leave good reviews will keep people coming and trying the food. In other words, the ratings are so compelling that every individual chooses to go to the restaurant. As a result, the portion of the population who goes reflects the overall population well. Hence the proportion of people who go and like it converges to the true proportion of people who like it. When $\alpha \in [0.5,\rho]$ we appear to have $\bar v_i\rightarrow \rho$. We interpret this to mean that when the quality is `so-so' (more accurately, when it is liked by a fair amount of the population), the average rating tends to be higher than the quality $\alpha$ itself. This suggests that a moderate quality restaurant can attain a degree of self-selectivity in its consumer base. Individuals are wary of the so-so rating and decide to trust their private signal. As a result, only individuals with a positive private signal decide to go to the restaurant. Among them, a larger fraction likes the restaurant than in the overall population, hence $\bar v_i>\alpha$. When $\alpha \in  [1-\rho, 0.5)$ it is harder to tell what happens: $\bar v_i$ converges to some value between $\alpha$ and $\rho$, increasing with respect to $\alpha$. It can be shown using Bayes' rule that when $\bar v_i$ converges, it approaches 
$$\operatorname{min}\left(\frac{\alpha \rho}{\alpha \rho+(1-\alpha) (1-\rho)}, \rho\right)$$
In other words, customer selection through private signals weighs the rating upwards for medium quality restaurants. The rating however is not bound to increase forever. If the rating becomes too large, everyone wants to go to the restaurant regardless of their private signal. It is the end of self-selection and everyone starts rating the business. The ratings become more and more reflective of the true business quality and start decreasing. Soon enough, the ratings become so-so again and only individuals with positive signals go to the venue. Intuitively, the process should lead the rating to converge hence the above mathematical expression. \midskip

\noindent Finally, we observe that some of the runs displayed in Figure 1 show a flat rating at around $0.2$. These correspond to runs where unlucky businesses were visited in early stages by negative customers. Their ratings subsequently took a hit, and the evidence showing that they are low-quality restaurants became overwhelming for anyone to knock on their door ever again.  \midskip

\noindent More formally, it has been noted in the `solution' section that if the rating falls below $1-\rho$, consumers will stop going to the restaurant regardless of their private signal. We refer to this as ``business death." So in a given simulation a business can die fairly quickly. Consider the following recursive formula for the rating:
$$\bar v_{i+1} = \left\{
     \begin{array}{lr}
       \bar v_i & \text{ if } i+1 \text{ doesn't go}\\
        \frac{|N_{i+1}|\bar v_i + v_{i+1}}{|N_{i+1}| + 1} & \text{ if } i+1 \text{ goes}
     \end{array}
   \right.$$
where $|N_{i+1}|$ is the number of consumers who went to the restaurant before $i+1$. If enough consumers who dislike the restaurant go, it may end up the case that $\bar v_i$ falls below $1-\rho$, in which case the business dies, because no consumer wants to go to a restaurant with $\bar v_i < 1-\rho$. We estimate the business death after $1000$ consumers by averaging over $1000$ simulated trials for over $5000$ $(\rho,\alpha)$ pairs. We reason that since $\bar v_i$ can only change very little at a time after $i > 1000$ that our estimate is a very good approximation for the total probability of business death. 
\begin{figure}[h!]
  \centering
    \includegraphics[width=0.7\textwidth]{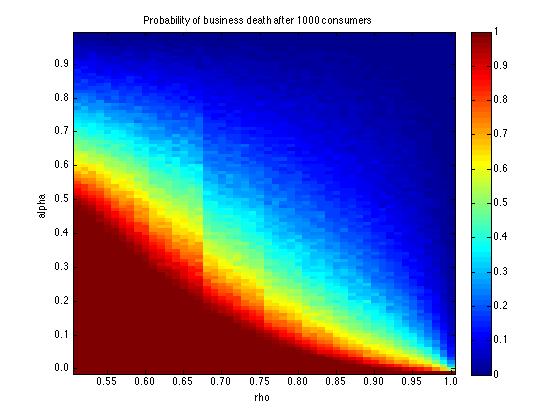}
  \caption{Probability of business death versus $(\alpha,\rho)$.}
\end{figure}

\noindent This figure has some nice properties. For a fixed value of $\rho$, the probability of business death decreases for increasing $\alpha$. When $\alpha = 0$, the restaurant fails no matter the value of $\rho$ because anyone who goes will leave a negative rating. Similarly when $\alpha = 1$, the restaurant cannot fail because anyone who goes loves it. When $\rho = 1$, everyone knows what they want, so any restaurant with $\alpha > 0$ can survive. Contrast this case with $\rho = 0.5$, when the private signal $x_i$ is essentially useless. In this case, only restaurants with $\alpha > 0.5$ can survive the rating system. More generally the worse individuals are at knowing what they like, the more likely they are to go to a restaurant they dislike, the more likely a restaurant is to be unlucky, receive the wrong customers and die. \midskip

\noindent One feature of the plot which is hard to explain is the vertical discontinuities at $\frac{2}{3}$ and $\frac{3}{4}$. The more refined we make this plot, the more pronounced those discontinuities become. The smooth transition between the regime where every restaurant dies and every restaurant survives is perhaps the most interesting feature of this graph. It is clear from this plot that if a restaurant got to decide $\alpha$ or $\rho$, they should prefer $\alpha = 1$, but if they have to have $\alpha < 1$ they want $\rho$ to be as large as possible. In other words however small $\alpha$ is, a business can avoid death by making sure that the right customers and only the right customers enter through the door. This effect can be sustained by maintaining an average rating and relying on customers to follow very accurate private signals. \midskip

\noindent Recognizing that the majority of consumers are operating within this paradigm, the keen (say level $k+1$) consumer would want an idea of how accurate the rating is to the true quality. For this reason we plot the \textit{rating bias}, defined as the average rating minus the quality $\alpha$.

\begin{figure}[h!]
  \centering
    \includegraphics[width=0.7\textwidth]{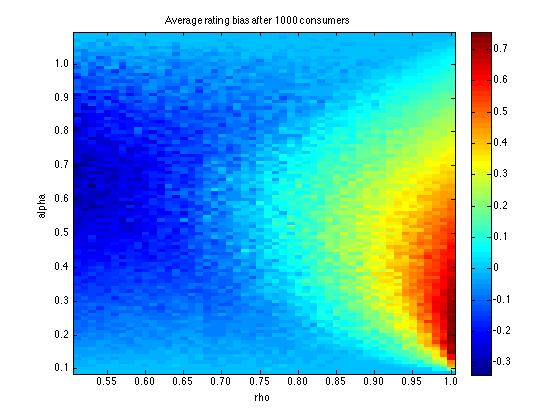}
  \caption{Rating bias $\bar v_{1000} - \alpha$ for various $(\rho,\alpha)$.}
\end{figure}

\noindent When $\alpha$ is very high or very low, the rating bias is nearly zero. This is a consequence of the fact that when everybody likes the restaurant, it will always get good ratings, and when nobody likes the restaurant, it will always get bad ratings. More generally, the rating bias seems to depend more on $\rho$ than on $\alpha$. When $\rho$ is large (so that everyone knows what they want), the lower quality restaurants get a high upward bias, i.e. their rating is higher than their quality. To an outsider who does not know what they want, in a world where everyone leaves reviews only for the places they like, the rating system is useless. Towards the periphery of the semi-circle on the right seems to be the `sweet spot' where the rating system favors restaurants, and conversely on the left semi-circle we have the rating system asserting a highly negative bias. Around where $\alpha = 0.5$ and $\rho = 0.5$ the downward bias makes it so restaurants have ratings significantly lower than their actual quality. Since business death is so prominent in this region of the plot, this is what probably kills their ratings.\midskip

\noindent We wanted to come up with one more mechanism to mitigate these rating biases, particularly the negative bias in moderate quality restaurants when people have low $\rho$. So in Figure 4 we plot the rating bias when only half of the population reads the reviews. The motivation for this is that business death is cause by customers reading bad reviews. In contrast, no business (at least for which $\alpha > 0$) can totally die when some consumers only follow their private signal.   

\newpage

\begin{figure}[h!]
  \centering
    \includegraphics[width=0.7\textwidth]{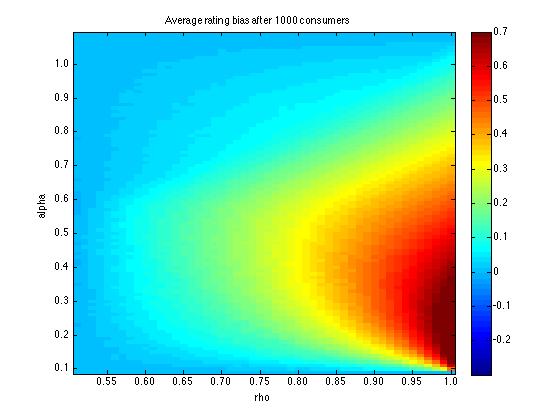}
  \caption{Rating bias $\bar v_{1000} - \alpha$ for various $(\rho,\alpha)$, $\frac{1}{2}$ readers}
\end{figure}

Introducing this sub-population sort of levelled out the bias in the average rating. To be more precise, the bias is almost entirely eliminated in the regime where $\rho < 0.75$. As hypothesized, the previously observed negative bias was due to business death, which has become less likely. In the regime where $\rho > 0.75$ the bias is still present, positive and large for small $\alpha$.\midskip

\noindent\textbf{Conclusions.} Through mathematical analysis and simulations, on-line ratings and their impact on businesses were characterized through two parameters: the inherent and objective restaurant quality factor, $\alpha$, and the accuracy of a customer's gut feeling about a business, $\rho$. In particular, it was shown that on-line ratings were seldom accurate mainly because of the low or high accuracy in customers' gut feelings.\midskip

\noindent When customers are very uncertain with regards to their preferences, they tend to purchase products or services that they do not like and leave bad reviews. If they so happen to be one of the first customers to rate the business, the bad review is a condemnation to death. On the other hand, accurate gut feelings tend to advantage medium-quality businesses through self-selection. In general, no matter how bad a business or product is, it will still be appreciated by a fraction of the population. By relying on accurate gut feelings, customers who decide to purchase the low-quality product will most often end up enjoying it and positively rating it. \midskip

\noindent Given that this effect is more pronounced for low quality businesses, it is very likely the case that on-line referral systems help low-quality businesses survive. Furthermore, if only a fraction of consumers read on-line reviews, the business death risk is reduced for low-quality businesses and on-line ratings further shift the scales in favor of low-quality businesses.\midskip

\noindent\textbf{Acknowledgements.} We'd like to thank our professor for his suggestions throughout this project, for always staying after class to chat about game theory and life, and for inspiring us to seed our random number generator with `JIM,' which yielded an altogether different picture of business death.

\begin{figure}[h!]
  \centering
    \includegraphics[width=0.45\textwidth]{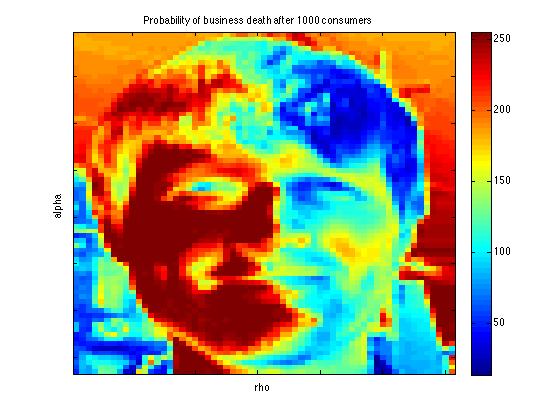}
  %\caption{Probability of business death versus $(\alpha,\rho)$.}
\end{figure}

\newpage

\noindent {\textbf{Relevant Code.}}
\begin{verbatim}
function[rate, death] = rating(a, p, N)
%% Given alpha-rho values a,p and a number of customers N,  
%% Returns the rating after N customers have opportunity to buy
    % Assume first person follows private signal
    rate = 0.5;
    t = 1;
    r = 0;%number of ratings so far
    
    % User never goes when alpha smaller than 1-rho
    while (rate > 1-p) && (t <= N)
        %User likes the restaurant with prob. alpha
        v = rand < a;
        %Message is correct with probability rho
        x = rand < (p*v + (1-p)*(1-v));
        % User always goes if alpha greater than rho 
        % or x = 1 and alpha greater than 1 - rho
        if (rate > p) || x
            rate = (r*rate + v) / (r + 1);
            r = r + 1;
        end
        t = t+1;
    end
    % indicates whether business died before 
    % the first N customers decided 
    death = (t < N);
end

%% Script to generate business death plots
alpha = 0:.01:1; I = length(alpha);
rho = 0.500001:.01:1; J = length(rho);
S = 10^3; N = 10^3; M = zeros(I,J);
for i=1:I
    for j=1:J
        p = rho(j); a = alpha(i); 
        B = zeros(1,N);
        for n=1:N
            [~,B(n)] = rating(a, p, S);
        end
        M(i,j) = sum(B)/N;
    end
end
imagesc(flipud(M));
\end{verbatim}

\vfill
\end{document}